\documentclass[aps,prc,preprint,groupedaddress]{revtex4}
\def\ohalf{{\textstyle{1\over 2}}}

\newcommand{\beq}{\begin{equation}}
\newcommand{\eeq}{\end{equation}}
\newcommand{\beqa}{\begin{eqnarray}}
\newcommand{\eeqa}{\end{eqnarray}}

\usepackage{epsfig}

\begin{document}

\title{The Role of 5-quark Components on the Nucleon Form Factors}

\author{Q. B. Li}
\email[]{ligb@pcu.helsinki.fi}
\affiliation{Helsinki Institute of Physics
POB 64, 00014 University of Helsinki, Finland}

\author{D. O. Riska}
\email[]{riska@pcu.helsinki.fi}
\affiliation{Helsinki Institute of Physics
and Department of Physical Sciences, POB 64,
00014 University of Helsinki, Finland}

\thispagestyle{empty}

\date{\today}
\begin{abstract}

The covariant quark model is shown to allow a phenomenological
description of the neutron electric form factor, $G_E^n(Q^2)$,
in the impulse approximation, provided that
the wave function contains minor ($\sim$ 3\%)
admixtures of the lowest energy sea-quark configurations.
While that form factor is not very sensitive to whether the $\bar q$ in the
$qqqq\bar q$ component is in the $P-$state or in
the $S-$state the calculated nucleon magnetic form factors are
much closer to the empirical values in the case of the
former configuration. In the case of the electric form factor
of the proton, $G_E^p(Q^2)$ a zero appears in the impulse
approximation close to 10 GeV$^2$, when the $\bar q$ is in the
$P-$state. That configuration, which may be interpreted as a pion loop
(``cloud'') fluctuation, also leads to a clearly
better description of the nucleon magnetic
moments. When the amplitude of the sea-quark admixtures
are set so as to describe the electric form factor of the neutron,
the $qqqq\bar q$ admixtures have the phenomenologically desirable
feature, that the electric form factor of the proton falls at a
more rapid rate with momentum transfer than the magnetic form factor.

\end{abstract}


\maketitle

\section{Introduction}

Phenomenological analyses of the electric form factor of the
neutron have indicated that the neutron may be viewed as a
3-quark core surrounded by a meson cloud \cite{Friedrich}.
This result is qualitatively consistent with hadronic model studies of the
electromagnetic and weak decay patterns of the $\Delta(1232)$
resonance, which show that the phenomenological failures of the
conventional $qqq$ model of the nucleon and the resonance may
be understood as a consequences of the lack of a meson cloud
component in the $\Delta(1232)$ \cite{satolee1,satolee2}. While
the covariant extensions of the $qqq$ quark model may be tuned to provide
a qualitatively satisfactory description of the extant data on the nucleon
form factors, including that on the electric form factor of the
neutron \cite{Bruno}, it is nevertheless natural to
extend the quark model to include explicit sea-quark
contributions \cite{An}, which more directly
may be interpreted as meson cloud
components. This is so much more the case as several measurements
of the $\bar d/\bar u$
asymmetry in the nucleon sea indicate the presence of
$qqqq\bar q$ components in the proton \cite{towell,NA51,NMC,HERMES}.

Here a calculation of the contribution of $qqqq\bar q$ components
in the nucleon wave functions to their electromagnetic form factors is
described in instant form kinematics. In view of the large number of possible
$qqqq\bar q$ configurations \cite{helminen},
the calculation is restricted to
those $qqqq\bar q$ configurations that are expected to have
the lowest energy with the $\bar q$ either in the excited
$P-$state or in the $S-$state. In the latter case positive
parity requires the $qqqq$ subsystem to be in the $P-$state. The
calculation is carried out with the covariant quark
model with instant form kinematics, which is similar to
the nonrelativistic quark model, save for the explicit
treatment of the constituent boosts.

The spatial wave function is taken to have a simple algebraic
form with two parameters, which are determined by fits to
the electric form factors of the nucleons. The amplitude of the
$qqqq\bar q$ component in the wave function is determined by
the fit to the electric form factor of the neutron. The
$qqqq\bar q$ contributions to the form factors do
depend on the radial part of the wave function, but with little
sensitivity to whether the
$\bar q$ is in the $S-$state or in the $P-$state.
In the case of the electric form factor of the proton, $G_E^p(Q^2)$,
the inclusion of the $qqqq\bar q$ component markedly improves the
description of the most recent data, which shows that the electric
form factor falls much faster with momentum transfer than
the corresponding magnetic form factor. In case where the antiquark
is in the $P-$state, the impulse approximation leads to
a zero in the calculated electric form factor between 10 and 11 GeV$^2$.
In this case the quantum numbers of the $qqqq\bar q$
configuration correspond to those of a pion loop fluctuation,
and hence admit an interpretation as a ``pion cloud''
configuration. This configuration is the preferred one
for the description of the nucleon magnetic moments.

The nucleon wave functions in the extended quark model
are described in section 2, along with the boosts to the
Breit frame that are needed in instant form kinematics. The calculation
of the electromagnetic form factors of the neutron and the proton is
described in section 3. Section 4 contains a summarizing
discussion.

\section{Nucleon wave functions with 3 and 5 quarks}

\subsection{The $qqq$ component}

The radial wave functions of the nucleons formed by n
constituents in their rest frame may be expressed
in terms of the following set of relative momenta:
\begin{equation}
\vec\xi_j ={1\over \sqrt{j+j^2}} \left
[\sum_{l=1}^j \vec k_l  - j \vec  k_{j+1} \right],
\quad  j=1,.., n-1\, .
\label{Jacobi}
\end{equation}
The rest frame is defined by the condition that
$\sum_{i=1}^n \vec k_i = 0$ .

In the case of the $qqq$ component, where all the constituent quarks are in
the ground state, the radial wave function will be taken to have the
form \cite{Bruno}:
\begin{equation}
\varphi(\vec\xi_1,\vec\xi_2)={\cal N}_3
{1\over (1+({\vec\xi_1\,^2+\vec\xi_2\,^2\over 2 b^2}))^a}\,.
\label{qqqR}
\end{equation}
Here ${\cal N}_3$ is a normalization constant.
The two parameters $a$ and $b$ may be determined by a fit to the
empirical proton form factor.

The spin-isospin part of the $qqq$ component shall be taken to have
the usual $SU(2)$ symmetric form, which is formed by the symmetric
combination of mixed symmetry spin and isospin wave functions:
$[3]_{FS}[21]_F[21]_S$ (Here $F$
stands for flavor and $S$ for spin). The complete wave function
for the nucleon may then be expressed in the compact form:
\begin{equation}
\psi_N(t,s)=\psi_{[111]}^C\,\varphi(\vec\xi_1,\vec\xi_2)
\sum_a C^{[3]}_{[21]_a,[21]_a}\psi^F_{[21]_a}(t) \,\psi^S_{[21]_a}(s)
\, .
\label{qqqW}
\end{equation}
Here the antisymmetric color component is denoted $\psi_{[111]}^C$ and the
mixed symmetry isospin and spin components are denoted
$\psi^F_{[21]_a}(t)$ and $ \psi^S_{[21]_a}(s)$ respectively. The $S_3$
Clebsch-Gordan coefficients $C^{[3]}_{[21]_a,[21]_a}$ take the
values $1/\sqrt{2}$ for the two mixed symmetry components $(112)$  ($a=1$)
and $(121)$ ($a=2$) respectively. The 3rd components of the
isospin of the proton and
the neutron are denoted $t$ and the corresponding spin components $s$.

\subsection{The $qqqq\bar q$ component}

The possible $qqqq\bar q$ components of the nucleons fall into 2
classes: those, in which the $qqqq$ subsystem is in the
$P-$state and the $\bar q$ is in the ground state, and those, in
which the $qqqq$ subsystem is in the ground state and the
$\bar q$ is in the $P-$state \cite{helminen}. A ``pion cloud''
configuration may be represented by the latter, as the antiquark
in the pseudoscalar pion has to be in a $P-$state. Which one of
the $qqqq\bar q$ configurations that has the lowest energy, and
consequently the largest amplitude in the proton, will depend
on the form of the hyperfine interaction between the quarks.

If the hyperfine interaction depends on spin and/or isospin, the
lowest energy configuration will be that with the most antisymmetric
spin and/or isospin symmetry, as those yield the largest
matrix elements of the operators $\sum_{i<j}\vec\sigma_i\cdot\vec\sigma_j$ ,
$\sum_{i<j}\vec\tau_i\cdot\vec\tau_j$ and
 $\sum_{i<j}\vec\tau_i\cdot\vec\tau_j\,\vec\sigma_i\cdot\vec\sigma_j$.
In the case, where the $\bar q$ is in the ground state, the lowest
energy configuration is that, in which the $qqqq$ subsystem has
the isospin-spin symmetry $[4]_{FS}[22]_F [22]_S$. When the
$\bar q$ is in the $P-$state the $qqqq$ subsystem with the mixed
flavor-spin symmetry  $[31]_{FS}[22]_F [31]_S$ has the lowest
energy.

\subsubsection{The antiquark in the $S-$state}

The $qqqq\bar q$ state with the lowest energy, for which the
$\bar q$ is in the ground state, has the $qqqq$ subsystem in
the $P-$state, with the mixed spatial symmetry $[31]_X$. The
wave function for this 5-quark system may be expressed in the
form \cite{zou1}:
\begin{eqnarray}
\psi_ N (t,s)^{5q} &=&
\sum_{a,b}\sum_{m,\bar s}
(1,\ohalf,m,\bar s\vert\, \ohalf,s)\,
C^{[1^{4}]}_{[31]_a[211]_a}\,C^{[4]}_{[22]_b [22]_b}
\nonumber\\
&&
\psi_{[211]_a}^C\,\varphi_{[31]_a ,m}(\{\vec\xi_i\})\,
\psi^F_{[22]_b}\,\psi^S_{[22]_b}\,
\bar \psi^C\,\bar\chi_{t,\bar s}\, .
\label{5qn}
\end{eqnarray}
Here the sum over $a$ runs over the 3 configurations of the $[211]_C$
and $[31]_X$ representations of $S_4$, and the sum over $b$ runs
over the 2 configurations of the $[22]$ representation of $S_4$
respectively~\cite{chen}. The $S_4$ Clebsch-Gordan coefficients
$C^{[1^{4}]}_{[31]_{a} [211]_{a}}$
take the values \cite{chen}:
\beq
C_{[31]_1 [211]_1}^{[1^4]}={1\over \sqrt{3}}\, ,\qquad
C_{[31]_2 [211]_2}^{[1^4]}={-1\over \sqrt{3}}\, ,\qquad
C_{[31]_3 [211]_3}^{[1^4]}={1\over \sqrt{3}}\, .
\eeq
The values of the $S_4$ Clebsch-Gordan coefficients
$C^{[4]}_{[22]_b [22]_b}$ are $1/\sqrt{2}$ for $b=1,2$. In
(\ref{5qn}) the color and spin-isospin wave functions
of the $\bar q$ are denoted $\bar\psi^C$ and
$\bar\chi_{t,\bar s}$ respectively.

The spatial wave functions
$\varphi_{[31]_a,m}$ for the complete $qqqq\bar q$
system (\ref{5qn}) will be products of symmetric function of the
Jacobi coordinates $\vec\xi_i$, $i=1,..4$ (\ref{Jacobi})
and one of the 3 vectors $\vec \xi_a$, $a=1,2,3$, if the
position coordinate of the $\bar q$
is taken to be $\vec r_5$.

\subsubsection{The antiquark in the $P-state$}

If the $\bar q$ antiquark is in the $P-$state, the
$qqqq$ subsystem is in the symmetric ground state. In this
case its spin-isospin state has to have the mixed symmetry
$[31]_{FS}$ in order to combine with the mixed symmetry
color state $[211]$ to a completely antisymmetric
state. The lowest energy spin-isospin configuration will
in this case be $[31]_{FS}[22]_F [31]_S$. The corresponding
complete wave
function may be written in the form:
\begin{eqnarray}
\tilde\psi_ N (t,s)^{5q} &=&
\sum_{a,b,c}\sum_{m,m',\bar s}
(1,\ohalf,m',\bar s\vert\, j,m) (j,1,m,S\vert\ohalf,s)\,
C^{[1^{4}]}_{[31]_a[211]_a}\,C^{[31]_a}_{[22]_b [31]_c}
\nonumber\\
&&
\psi_{[211]_a}^C\,\varphi_{[4],m'}(\{\vec\xi_i\})\,
\psi^F_{[22]_b}\,\psi^S_{[31]_c(S)}\,
\bar\psi^C\,\bar\chi_{t,\bar s}\, .
\label{5qnb}
\end{eqnarray}
In this case the spatial wave function $\varphi_{[4],m}$
will be a symmetric function of the Jacobi coordinates
$\vec \xi_i$, $i=1,..4$ multiplied by $\vec\xi_{4,m}$ (\ref{Jacobi}).

In (\ref{5qnb}) the 3rd component of the spin of the
$qqqq$ subsystem takes the values 1,0 and -1. In the case of
the elastic nucleon form factors contributions may arise from
the terms with $j=1/2$ and $j=3/2$.

\subsubsection{The spatial wave function models}

The spatial wave function model for the $qqqq\bar q$ state, for which the
spatial wave function of the $qqqq$ subsystem has the
mixed symmetry $[31]_X$ the following algebraic form will be
employed
\begin{equation}
\varphi_{[31]_a ,m}(\{\vec\xi_i\}) = {\cal N}_{[31]}{\xi_{a,m}\over
(1+{(\sum_{j=1}^4\vec \xi_j\,^2)\over 2 B^2})^{(A+1)}}\, , a=1,2,3 \,.
\label{31sp}
\end{equation}
Here ${\cal N}_{[31]}$ is a normalization constant.
This wave function has the appropriate threshold behavior for a
$P-$state wave function \cite{Bruno}.
The analogous wave function for the $qqqq\bar q$ state, for which
the spatial wave function of the $qqqq$ subsystem is symmetric,
and where the $\bar q$ is in the $P-$state, is taken to be
\begin{equation}
\varphi_{[4],m}(\{\vec\xi_i\}) = {\cal N}_{[4]}{\xi_{4,m} \over
(1+{(\sum_{j=1}^4\vec \xi_j\,^2)\over 2 B^2})^{(A+1)}}\,.
\label{4sp}
\end{equation}
The symmetric function of Jacobi coordinates in eqs. (\ref{31sp})
and (\ref{4sp})
leads to the fact that ${\cal N}_{[4]}={\cal N}_{[31]}$.

\subsection{Breit frame wave functions}

In a covariant calculation the elastic nucleon form factors
in instant form kinematics are most conveniently calculated
as matrix elements of the initial and final wave functions
in the Breit frame. For this the rest frame wave functions
described above have to be boosted to the Breit frame, in which
there is zero energy transfer to the nucleon.

Without loss of generality the momentum transfer to the nucleon
may be taken to define the $z-$axis. A generalization of the
instant form boost relations of ref. \cite{Bruno}
then leads to the following relations between the
constituent momenta
$\vec p_i$ in the Breit frame and in the rest frame (primes
denote final state momenta):
\begin{eqnarray}
&&\vec  p_{i\perp} = \vec k_{i\perp}
 = \vec k_{i\perp}\,^{'} = \vec{ p}_{i \perp}\,^{'}\, ,
\nonumber \\
&&
 p_{i\parallel} = v_{0} k_{i\parallel}+v_{\parallel}\,
\omega_i \, ,  \nonumber \\
&&
 p_{i\parallel}^{'} = v_{0}^{'} k_{i\parallel}^{'}
+v_{\parallel}^{'}\, \omega_i^{'} \, ,  \nonumber \\
&&E_i= v_{\parallel}k_{i\parallel}+v_{0}\omega_i\, ,
\nonumber\\
&&E_i^{'}=v_{\parallel}^{'}k_{i\parallel}^{'}
+v_{0}^{'}\omega_i^{'}\, .
\label{boosts}
\end{eqnarray}
Here the energy components are defined as
\begin{eqnarray}
&&\omega_i=\sqrt{\vec k_i\,^2 + m^2}\,,\quad
\omega_i^{'}=\sqrt{\vec k_i\,^{'2} + m^2}\, ,\nonumber\\
&&E_i=\sqrt{\vec p_i\,^2 + m^2}\, ,\quad
E_i^{'}=\sqrt{\vec p_i\,^{'2} + m^2}\, .
\label{energies}
\end{eqnarray}
In these relations $m$ denotes the constituent mass
and $v=\{v_{0},\vec 0_\perp,v_{\parallel}\}$ and
$v'=\{v_{0}^{'},\vec 0_\perp,v_{\parallel}^{'}\}$ the
constituent boost velocities in the initial and final
states. These satisfy the constraint $v^2=v^{'2}=-1$.

In instant form kinematics the boost velocities may be
defined as \cite{Bruno}:
\begin{eqnarray}
&&v_{\parallel}=-{Q\over 2\sum_{i=1}^n \omega_i}\, ,
\nonumber\\
&&v_{\parallel}^{'}=\,\,\,\,{Q\over 2\sum_{i=1}^n \omega_i^{'}}\, .
\label{Vboosts}
\end{eqnarray}

In the calculation of the nucleon form factors the Jacobian
matrix of the transformations (\ref{boosts}) are also required.
In the present application the electromagnetic coupling
shall be assumed to take place on only one of the constituents.
If this is taken to be first of the constituents,
these Jacobians take the following form for the systems of
3 and 5 constituents:
\begin{eqnarray}
J_3&=&{\omega_2\omega_3\over E_2 E_3}(1-v_{\parallel}
{k_{1\parallel}\over E_1})\, ,
\label{Jac3}\\
J_5&=&{\omega_2\omega_3\omega_4\omega_5
\over E_2 E_3 E_4 E_5}(1-v_{\parallel}
{k_{1\parallel}\over E_1})\, ,
\label{Jac5}
\end{eqnarray}
for the initial state coordinates. The corresponding
expressions for the final state coordinates are obtained
by replacement of the variables by the corresponding primed variables.

As the spin quantization axis is rotated by the boosts, there is
in principle a need to take this into account by appropriate
Wigner rotations of the spin variables. The numerical
significance of the Wigner rotations has, however, been found
to be but minor in the region of momentum transfer
up to 10 GeV$^2$, which is relevant for form factors
\cite{coeris,Bruno}. The Wigner rotations will for reasons
of simplicity therefore not be considered here .

\section{The nucleon form factors}

\subsection{Definitions}

The electric and the magnetic nucleon form factors may be
defined as the following matrix elements of the
electromagnetic current density operator \cite{coekei}:
\begin{eqnarray}
G_E(Q^2)&=&\sqrt{1+\tau}\,\langle\, \ohalf \vert\,
J_0\,\vert\, \ohalf\,\rangle \, ,
\label{Eff}
\\
G_M(Q^2)&=&{\sqrt{1+\tau}\over \sqrt{\tau}}\,\langle \,\ohalf\, \vert\,
J_x \,\vert -\ohalf\,\rangle \, .
\label{Mff}
\end{eqnarray}
Here $\eta$ is defined as
\begin{equation}
\tau = {Q^2\over 4 M^2} \, ,
\label{eta}
\end{equation}
where $M$ is the nucleon mass.

\begin{figure}[t]
\vspace{20pt}
\begin{center}
\mbox{\epsfig{file=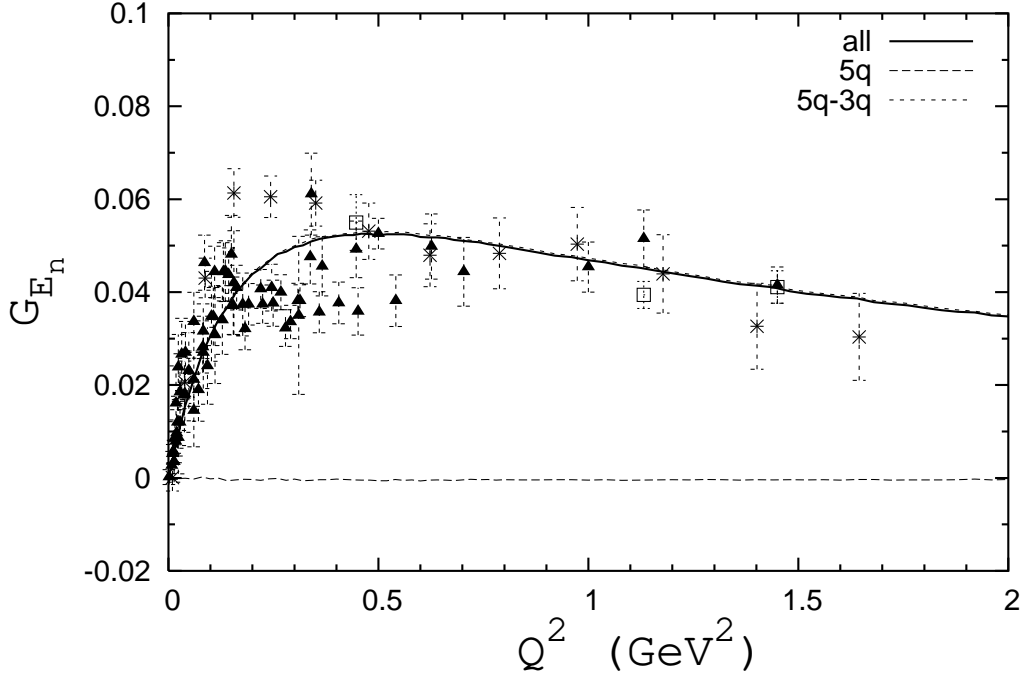, width=140mm}}
\caption{Calculated neutron electric form factor in the
presence of a $qqqq\bar q$ contribution with
the antiquark in the $S-$state . The long-dash curve
represents the diagonal  contribution from a $qqqq\bar q$
component with
3 \% probability, and the
short-dash curve the corresponding contribution from the
$qqq-qqqq\bar q$
transition matrix elements. The solid curve shows the
combined result. The data points are from
ref. \cite{genschi} (double crosses) \cite{madey} (boxes) and
\cite{genA1,plaster} and references therein (triangles).}
\label{genS}
\end{center}
\vspace{10pt}
\end{figure}

The form factors are obtained as integrals over the momenta
in the Breit frame. As the wave functions depend on rest frame
momenta the rest frame each wave function for the $qqq$ and
$qqqq\bar q$ subsystem in the integrand has to be multiplied by the
square root of the appropriate Jacobian - $J_3$ (\ref{Jac3}) or
$J_5$ (\ref{Jac5}) for initial states or the corresponding ones with primed
coordinates for the final states respectively.

The current operator of a constituent quark has the following
matrix elements:
\begin{eqnarray}
\langle\, \ohalf \vert\,
J_0\,\vert\, \ohalf\,\rangle \,& = &\sqrt{{(E'+m)(E+m)\over
4 E' E}}\bigg\{1+{\vec p\,'\cdot \vec p\over (E'+m)(E+m)}
\bigg\}\, ,
\\
\langle \,\ohalf\, \vert\,
J_x \,\vert -\ohalf\,\rangle \, &=& {1\over 2}
\sqrt{{(E'+m)(E+m)\over
4 E' E}} \bigg\{ { \vert Q\vert (E'+E+2m)\over (E'+m)(E+m)}
\nonumber\\
&&-{(p_{\parallel}+p_{\parallel}\,^{'})
(E'-E)\over (E'+m)(E+m)} \bigg\} \, .
\label{elasticJ}
\end{eqnarray}
The transition current operator for the transition
$q\bar q\rightarrow \gamma$ has the corresponding
matrix elements:
\begin{eqnarray}
\langle \,\ohalf\, \vert\,
J_0^{(a)} \,\vert \ohalf\,\rangle \, &=& {1\over 2}
\sqrt{{(E'+m)(E+m)\over
4 E' E}} \bigg\{ { \vert Q\vert (E'+E+2m)\over (E'+m)(E+m)}
\nonumber\\
&&+{(p_{\parallel}-p_{\parallel}\,^{'})
(E'-E)\over (E'+m)(E+m)}\bigg\}\, ,
\\
\langle\, \ohalf \vert\,
J_x^{(a)}\,\vert\, -\ohalf\,\rangle \, &=& \sqrt{{(E'+m)(E+m)\over
4 E' E}}\bigg\{1-{p_{\parallel}p_{\parallel}^{'}\over (E'+m)(E+m)}
\bigg\}\, .
\label{transitionJ}
\end{eqnarray}

\subsection{The electric form factors}

\subsubsection{The electric form factor of the neutron $G_E^n$}

Consider first the electric form factor of the neutron $G_E^n$. As this
form factor, with the present $SU(2)$ symmetric model wave function,
takes no contribution at all from the diagonal transitions between the
$qqq$ component, the experimental values may be employed to set the
amplitude of the $qqqq\bar q$ component in the neutron as well
as the parameters in
the wave function model phenomenologically.

\begin{figure}[t]
\vspace{20pt}
\begin{center}
\mbox{\epsfig{file=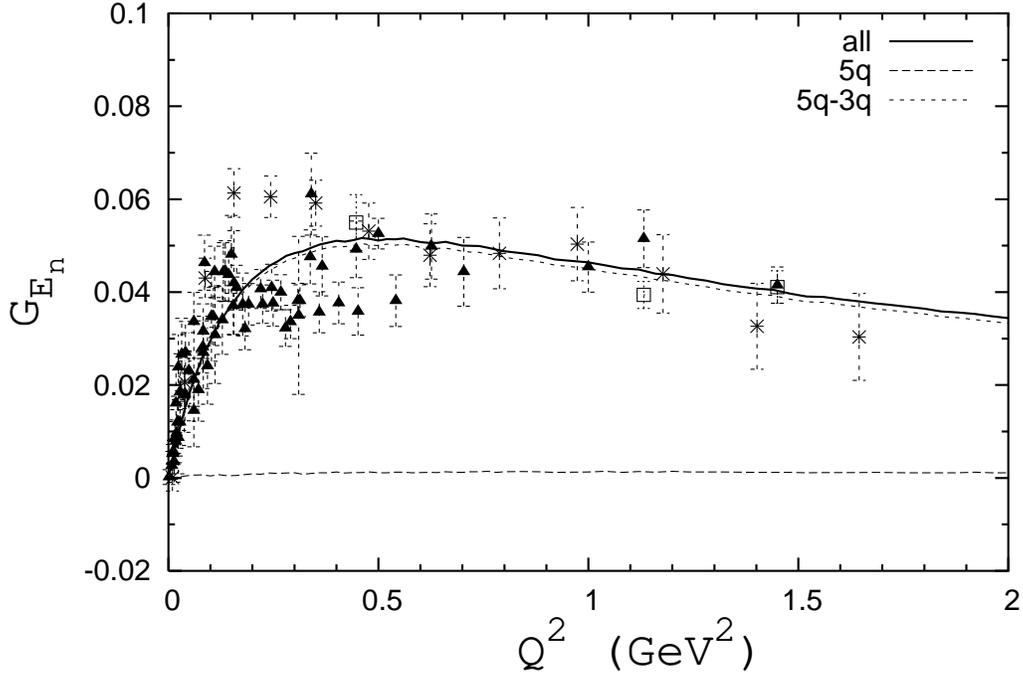, width=140mm}}
\caption{Calculated neutron electric form factor in the
presence of a $qqqq\bar q$ contribution with
the antiquark in the $P-$state . The long-dash curve
represents the diagonal  contribution from a $qqqq\bar q$
component with
3 \% probability, and the
short-dash curve the corresponding contribution from the
$qqq-qqqq\bar q$
transition matrix elements. The solid curve shows the
combined result. The data points are the same
as in Fig.\ref{genS}.}
\label{genP}
\end{center}
\vspace{10pt}
\end{figure}

The calculated neutron electric
form factors are shown in Figs. \ref{genS}
and \ref{genP} as obtained with a 3\% probability for the
$qqqq\bar q$ component, for both the cases where the
$\bar q$ is in the $S-$state and in the $P-$state,
respectively. The wave function parameters employed
in these calculations
are listed in Table \ref{parametersN}.

\begin{table}
\caption{The parameters of the nucleon
wave function components.\label{parametersN}}
\begin{ruledtabular}
\begin{tabular}{lcccc}
$m_q$ (MeV)  & $b_3$ (MeV) & $b_5$ &$a_3$ &$a_5$ \\
\hline

  260       &  310        &   315 & 3.9 & 3.0  \\
\end{tabular}
\end{ruledtabular}
\end{table}

These results reveal that the net contribution from the diagonal
matrix elements
of the $qqqq\bar q$ components in both cases is insignificantly small.
The notable contribution arises from the transition matrix elements
between the $qqqq\bar q$
and the $qqq$ components. The results also show
that the there is little discrimination between the cases, where the
antiquark
is in the $S-$ or in the $P-$states.

\subsubsection{The electric form factor of the proton $G_E^p$}

\begin{figure}[t]
\vspace{20pt}
\begin{center}
\mbox{\epsfig{file=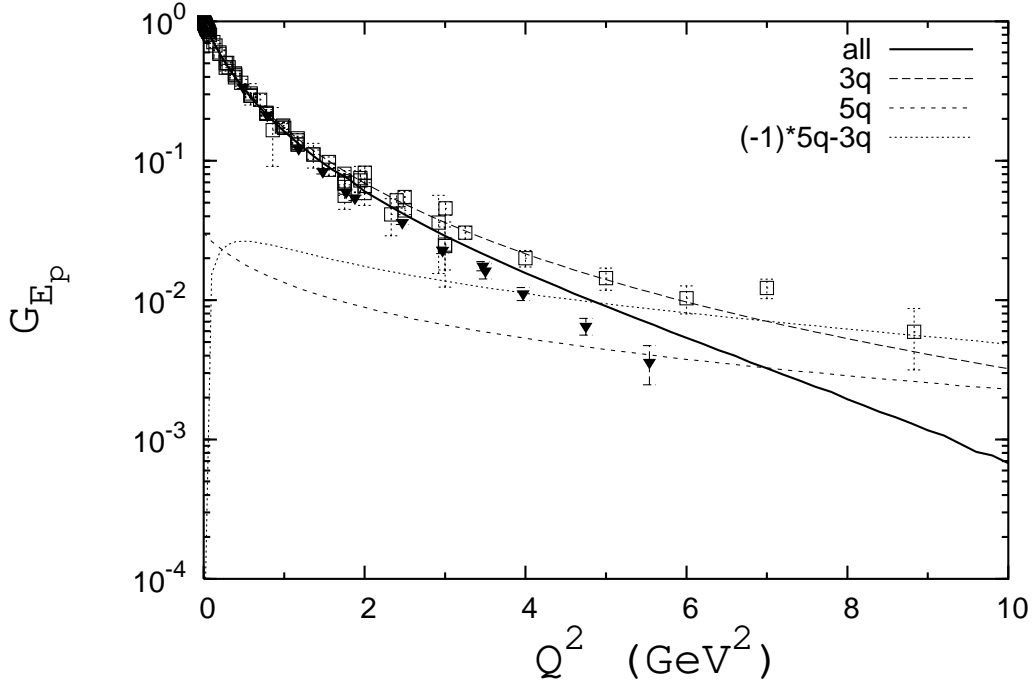, width=140mm}}
\caption{Calculated proton electric form factor in the
presence of a $qqqq\bar q$ contribution with
the antiquark in the $S-$state. The long-dash curve
represents the diagonal  contribution from
the $qqq$ component, and the short-dash curve
that from the diagonal  $qqqq\bar q$
component matrix elements with
3 \% probability. The dotted curve represents the
corresponding contribution from the
$qqq-qqqq\bar q$
transition matrix elements. The solid curve shows the
combined result. The data points are from
ref. \cite{jlab} and references therein.}
\label{gepS}
\end{center}
\vspace{10pt}
\end{figure}

\begin{figure}[t]
\vspace{20pt}
\begin{center}
\mbox{\epsfig{file=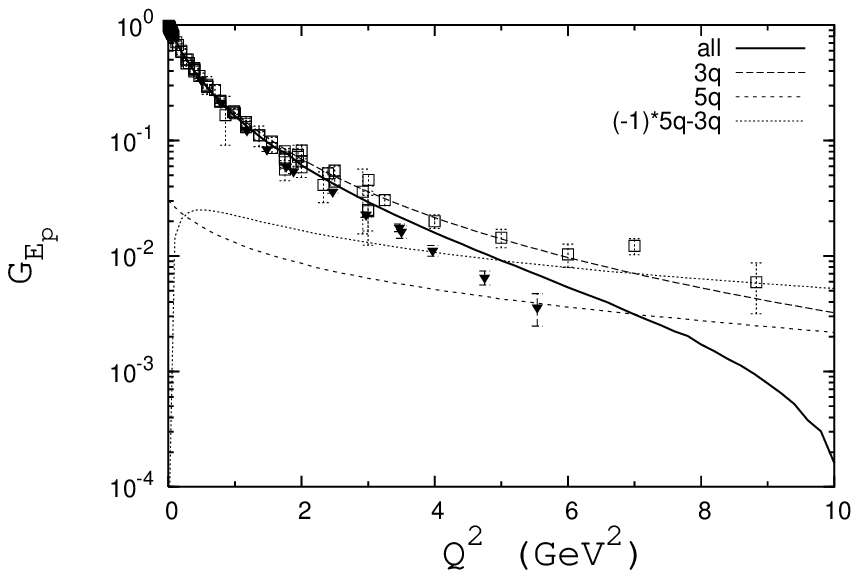, width=140mm}}
\caption{Calculated proton electric form factor in the
presence of a $qqqq\bar q$ contribution with
the antiquark in the $P-$state.
The long-dash curve
represents the diagonal  contribution from
the $qqq$ component, and the short-dash curve
that from the diagonal  $qqqq\bar q$
component matrix elements with
3 \% probability. The dotted curve represents the
corresponding contribution from the
$qqq-qqqq\bar q$
transition matrix elements. The solid curve shows the
combined result. The data points are the same
as in Fig.\ref{gepS}.}
\label{gepP}
\end{center}
\vspace{10pt}
\end{figure}

The calculated proton electric form factors are shown in in Figs. \ref{gepS}
and \ref{gepP}, again with a 3\% probability for the
$qqqq\bar q$ components for both the cases, where the
$\bar q$ is in the $S-$state and in the $P-$state respectively.

The results in Figs. \ref{gepS} and \ref{gepP} show that
the contribution from the diagonal matrix elements
of the $qqqq\bar q$ component are insignificant
in comparison to that from the main $qqq$ component.
The $qqq-qqqq\bar q$ transition matrix elements are,
however, significant, the significance growing with
momentum transfer.

The sign of the transition matrix
element contribution, which depends on the
relative sign of the $qqq$ and $qqqq\bar q$
wave function components, is determined here
by the sign of the contribution to
$G_E^n$. The effect of including the $qqqq\bar q$ component
on the calculated values for $G_E^p$ leads to a much
better description of the empirical form factor that
is determined by polarization transfer, than what is
possible without such a component \cite{Bruno}.
The role of the $qqqq\bar q$ component is to bring about
the desired faster falloff with $Q^2$ of $G_E^p$ than
of $G_M^p$. In the case of the conventional $qqq$
quark model, the momentum dependence of the electric
and magnetic form factors are very similar unless
explicit interaction current contributions are
considered.

In the case when the $\bar q$ is in the $P-$state,
the calculated electric form factor will have a zero between
10 and 11 GeV$^2$ in the impulse approximation. Such
a zero cannot be achieved in instant form kinematics
in the impulse approximation without additional
wave function components beyond the basic
$qqq$ component. A similar zero does, however,
appear naturally in the impulse approximation
in front form kinematics already in the $qqq$
model \cite{Bruno}.

\begin{figure}[t]
\vspace{20pt}
\begin{center}
\mbox{\epsfig{file=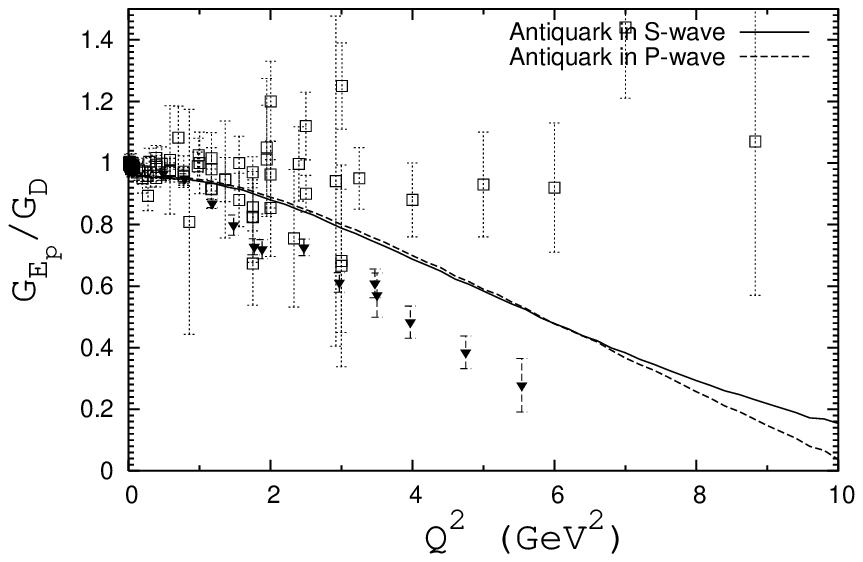, width=140mm}}
\caption{Ratios of the calculated proton electric 
form factors to the dipole form in the
presence of a $qqqq\bar q$ contribution with
the antiquark in the $S-$ (solid curve) and $P-$states
(dashed curve).
The data points correspond to those in Fig.\ref{gepS}.}
\label{gepD}
\end{center}
\vspace{10pt}
\end{figure}

In Fig.\ref{gepD} a plot of the ratios of the
calculated electric form factors of the proton
to the phenomenological dipole form is given.
These curves show that the two wave function configurations
lead to different results only above 7 GeV$^2$.
Both curves fall close to the values that
are extracted from the empirical cross section
by the forward-backward separation method, once
the correction from two-photon exchange has been
accounted for \cite{blunden,afa}.

The electric mean square radii that correspond to these form factors
are listed in Table \ref{electricradius}. In the case of the
proton mean square radius the main contribution is that from
the $qqq$ component in the wave function. The transition
matrix elements yield contributions that are smaller by an
order of magnitude, while the diagonal matrix elements of the
$qqqq\bar q$ components are insignificant. 

In the case of the
neutron, the main contribution arises from the transition
matrix elements between the $qqq$ and the $qqqq\bar q$ components.
The net value for the mean square radius is very close to
the empirical value, in the case where the antiquark is in the
$P-$state.

\begin{table}
\caption{The calculated nucleon
electric radii.\label{electricradius}}
\begin{ruledtabular}
\begin{tabular}{lccccc}
&&Proton \quad\quad$\langle r_e^2\rangle_{exp}= 0.774 \quad fm^2$&&
\\
\hline
$\bar q$ state  & $qqq$ & $qqqq\bar q$ & $qqqq\bar q \rightarrow qqq$
 &Total \\
\hline

$S$ &  0.675       &  0.009       &   0.052   & 0.736 \\
$P$ &  0.675       &  0.005       &   0.049   & 0.730 \\
\hline
&&Neutron \quad\quad$\langle r_e^2\rangle_{exp}= -0.113 \quad fm^2 $&&
\\ \hline

$S$ &  0.    &  -0.005     &  -0.103 & -0.108  \\
$P$ &  0.    &  -0.012     &  -0.098 & -0.110 \\
\end{tabular}
\end{ruledtabular}
\end{table}

\subsection{The magnetic form factors}

\subsubsection{The magnetic form factor of the proton $G_M^p$}

\begin{figure}[t]
\vspace{20pt}
\begin{center}
\mbox{\epsfig{file=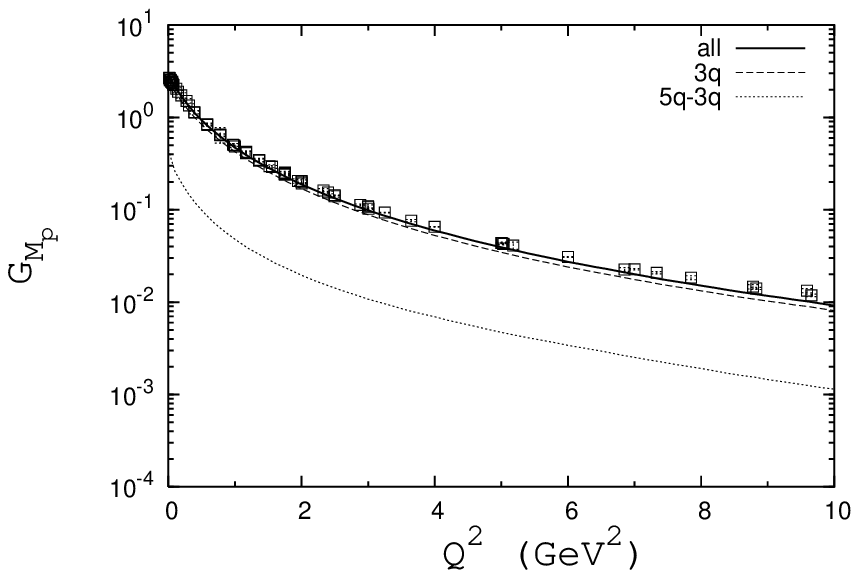, width=140mm}}
\caption{Calculated proton magnetic form factor in the
presence of a $qqqq\bar q$ contribution with
the antiquark in the $S-$state. The long-dash curve
represents the diagonal  contribution from
the $qqq$ component. The diagonal  $qqqq\bar q$
component matrix elements with
3 \% probability is insignificant and therefore
not shown. The dotted curve represents the
corresponding contribution from the
$qqq-qqqq\bar q$
transition matrix elements. The solid curve shows the
combined result. The data points are from
ref. \cite{jlab} and references therein.}
\label{gmpS}
\end{center}
\vspace{10pt}
\end{figure}

\begin{figure}[t]
\vspace{20pt}
\begin{center}
\mbox{\epsfig{file=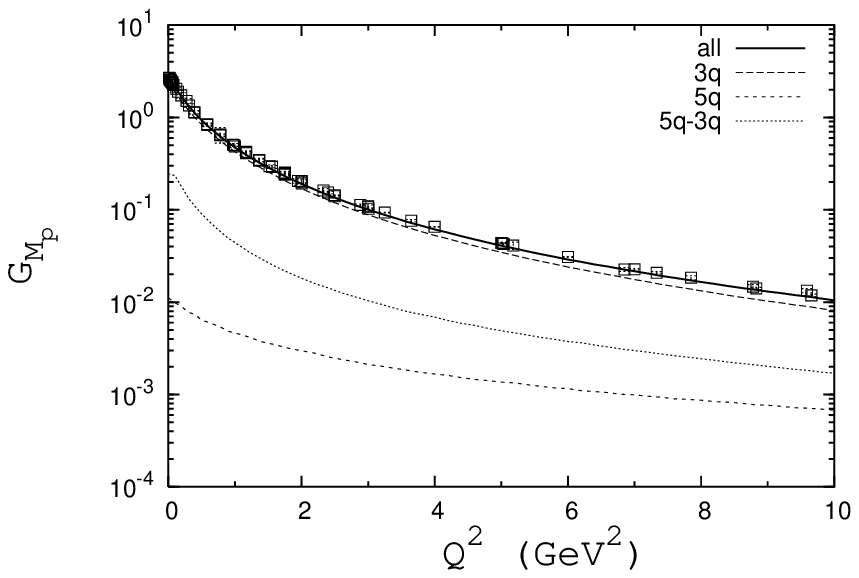, width=140mm}}
\caption{Calculated proton magnetic form factor in the
presence of a $qqqq\bar q$ contribution with
the antiquark in the $P-$state. The long-dash curve
represents the diagonal  contribution from
the $qqq$ component, and the short-dash curve
that from the diagonal  $qqqq\bar q$
component matrix elements with
3 \% probability. The dotted curve represents the
corresponding contribution from the
$qqq-qqqq\bar q$
transition matrix elements. The solid curve shows the
combined result. The data points are from
ref. \cite{mmd} and references therein.}
\label{gmpP}
\end{center}
\vspace{10pt}
\end{figure}

The calculated magnetic form factors of the proton
are shown in Figs. \ref{gmpS} and \ref{gmpP}.
In the case where the $\bar q$ is in the $S-$state
there is no diagonal contribution from the
$qqqq\bar q$ component.
In the
case, where the $\bar q$ is in the $P-$state
the contribution from the diagonal matrix
elements of the $qqqq\bar q$ component is smaller
by two orders of magnitude than that from the
corresponding $qqq$ matrix elements. The contribution
from the $qqqq\bar q$ transition matrix elements is
in both cases smaller by one order of magnitude
than that from the diagonal matrix elements of
the $qqq$ components. The overall effect of the
$qqqq\bar q$ components is small.

The corresponding magnetic moments are listed in
Table \ref{magmoments}. While the diagonal
contributions of the $qqqq\bar q$ components to these
are insignificant, the transition matrix elements
are substantial. As the magnetic moment contribution
to the proton magnetic moment from the $qqqq\bar q$
component is smaller in the case, where the $\bar q$
is in the $P-$state, that configuration is preferred.
This configuration is the one, which admits an interpretation
as a pion loop fluctuation.

\begin{figure}[t]
\vspace{20pt}
\begin{center}
\mbox{\epsfig{file=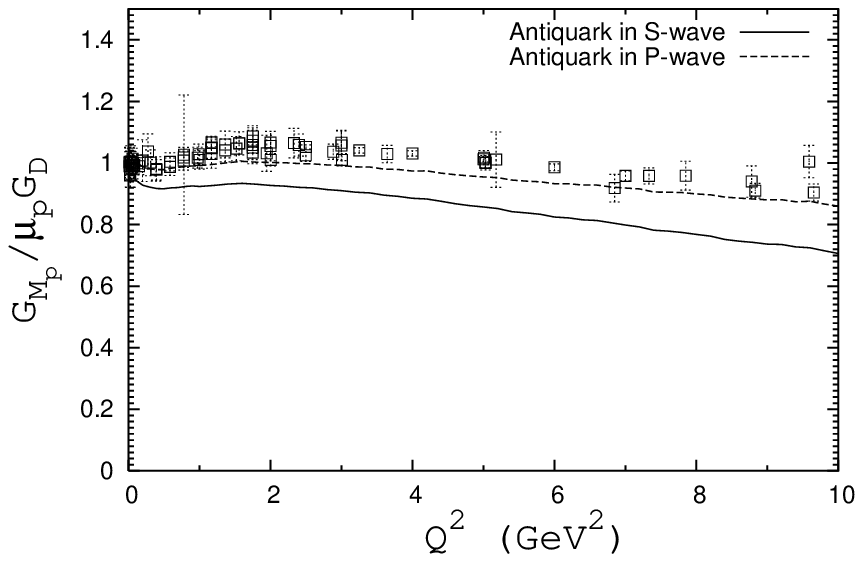, width=140mm}}
\caption{Ratios of the calculated proton magnetic 
form factors to the dipole form in the
presence of a $qqqq\bar q$ contribution with
the antiquark in the $S-$ (solid curve) and $P-$states
(dashed curve).
The data points correspond to those in 
ref. \cite{mmd} and references therein.}
\label{gmpD}
\end{center}
\vspace{10pt}
\end{figure}

In Fig.\ref{gmpD} a plot of the ratios of the
calculated magnetic form factors of the proton
to the phenomenological dipole form is given.
These curves reveal a clear phenomenological
preference for the wave function configuration,
in which the antiquark is in the $P-$state.
In that configuration the calculated magnetic
form factor follows the empirical one
fairly well over the whole range of momenta
considered. 

\subsubsection{The magnetic form factor of the neutron $G_M^n$}

\begin{table}
\caption{The calculated nucleon
magnetic moments.\label{magmoments}}
\begin{ruledtabular}
\begin{tabular}{lccccc}
&&Proton \quad\quad$\mu_{exp}= 2.79 \mu_N$&&
\\
\hline
$\bar q$ state  & $qqq$ & $qqqq\bar q$ & $qqqq\bar q \rightarrow qqq$
 &Total \\
\hline

$S$ &  2.51       &  0.0       &   0.45 & 2.97 \\
$P$ &  2.51 & 0.01 &  0.25                       & 2.77 \\
\hline
&&Neutron \quad\quad$\mu_{exp}=-1.91 \mu_N$&&
\\ \hline

$S$ &  -1.67       &  0.01       &  - 0.90 & -2.56 \\
$P$ &  -1.67 & -0.02 &  -0.49                       & -2.18 \\
\end{tabular}
\end{ruledtabular}
\end{table}

\begin{figure}[t]
\vspace{20pt}
\begin{center}
\mbox{\epsfig{file=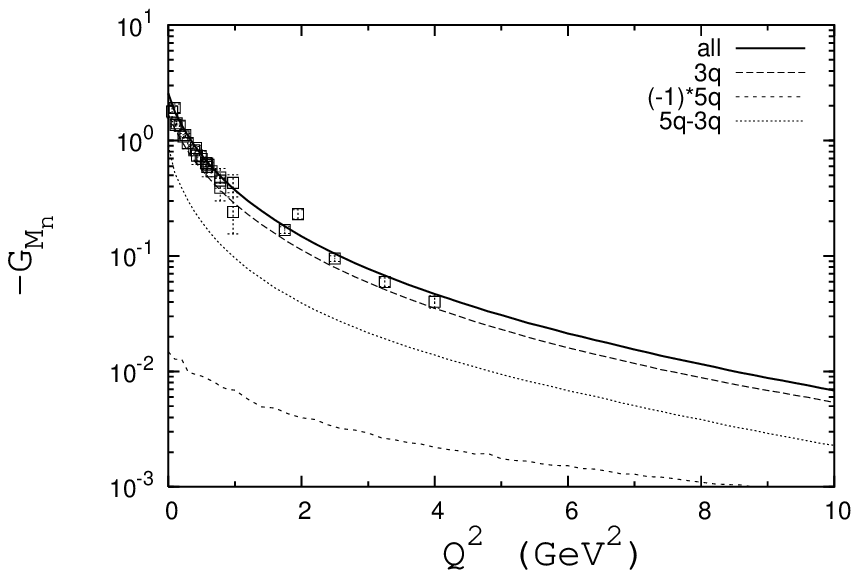, width=140mm}}
\caption{Calculated neutron magnetic form factor in the
presence of a $qqqq\bar q$ contribution with
the antiquark in the $S-$state. The long-dash curve
represents the diagonal  contribution from
the $qqq$ component, and the short-dash curve
that from the diagonal  $qqqq\bar q$
component matrix elements with
3 \% probability. The dotted curve represents the
corresponding contribution from the
$qqq-qqqq\bar q$
transition matrix elements. The solid curve shows the
combined result. The data points are from
ref. \cite{mmd,jlabx,plaster} and references therein.}
\label{gmnS}
\end{center}
\vspace{10pt}
\end{figure}

\begin{figure}[t]
\vspace{20pt}
\begin{center}
\mbox{\epsfig{file=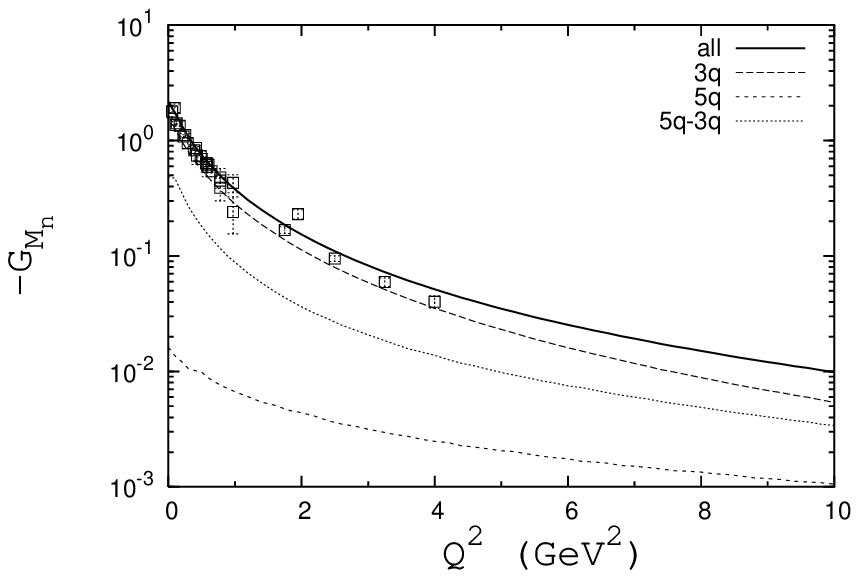, width=140mm}}
\caption{Calculated neutron magnetic form factor in the
presence of a $qqqq\bar q$ contribution with
the antiquark in the $P-$state. The long-dash curve
represents the diagonal  contribution from
the $qqq$ component, and the short-dash curve
that from the diagonal  $qqqq\bar q$
component matrix elements with
3 \% probability. The dotted curve represents the
corresponding contribution from the
$qqq-qqqq\bar q$
transition matrix elements. The solid curve shows the
combined result. The data points are from
ref. \cite{mmd,jlabx,plaster} and references therein.}
\label{gmnP}
\end{center}
\vspace{10pt}
\end{figure}

The calculated magnetic form factors of the neutron
are shown in Figs. \ref{gmnS} and \ref{gmnP}.
In this case the sign of the (small) diagonal $qqqq\bar q$
contribution depends on whether or not the $\bar q$
is in the $S-$ or in the $P-$state. The rate of
falloff with momentum transfer is somewhat faster
in the former case.

The non-diagonal contribution from the $qqqq\bar q - qqq$ transition
matrix elements to the neutron magnetic moments that are given in
Table \ref{magmoments} are about twice as large in the case of the
neutron as in the case of the proton. In this case there is a clear
preference for the $qqqq\bar q$ configuration, in which the $\bar q$
is in the $P-$state over that in which the $\bar q$ is in the
$S-$state. The large value for the magnetic moment in the latter
case agrees with the corresponding value found in ref. \cite{An} in
the harmonic oscillator model when the spatial extents of the $qqq$
and $qqqq\bar q$ configurations are equal.

While the value of the calculated magnetic moment of the proton
is very close to the empirical value if the antiquark
in the $qqqq\bar q$ component is in the $P-$state, the corresponding
value of the magnetic moment of the neutron is too large by
14 \%. This indicates that the model wave functions employed
for the $qqqq\bar q$ are too crude to describe both the form
factors and the static observables simultaneously. A reduction
of the probability of the $qqqq\bar q$ component from 3\% to
2\% would reduce the overestimate by half, but at the
price of a slighly less true description of the
electric form factor of the neutron.  

The calculated electric mean square radii of the neutron are
listed in Table \ref{electricradius}. In this case the
calculated values for both the wave function models are
similar, which is to be expected, as the wave function
parameters were chosen so that the calculated electric
form factor of the neutron would follow the empirical
shape.

\begin{figure}[t]
\vspace{20pt}
\begin{center}
\mbox{\epsfig{file=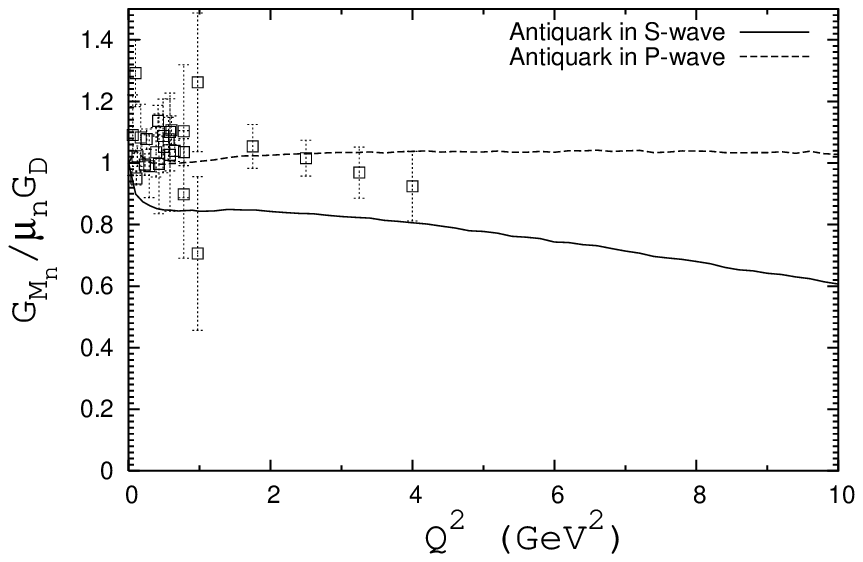, width=140mm}}
\caption{Ratios of the calculated neutron magnetic 
form factors to the dipole form in the
presence of a $qqqq\bar q$ contribution with
the antiquark in the $S-$ (solid curve) and $P-$states
(dashed curve).
The data points correspond to those in
ref. \cite{mmd,jlabx,plaster} and references therein.}
\label{gmnD}
\end{center}
\vspace{10pt}
\end{figure}

In Fig.\ref{gmnD} a plot of the ratios of the
calculated magnetic form factors of the neutron
to the phenomenological dipole form is given.
These curves again show that the wave function 
configuration,
in which the antiquark is in the $P-$state,
leads to better agreement with the
the empirical one
over the range of momenta
considered.

\section{Discussion}

The present exploratory study of the possibility to extend the
the $SU(2)$ constituent quark model so that it provides a
qualitative description of all the 4 electromagnetic form
factors of the nucleon reveals that this is possible once
a minor $qqqq\bar q$ component is included in the wave
function. The results also indicate a clear preference for
a $qqqq\bar q$ configuration, which has the same quantum
number as a pion loop fluctuation (``pion cloud''). In the
present work, the spatial extent of the $qqqq\bar q$ component
were taken to be similar to that of the main $qqq$
component, even though a pion cloud configuration might in contrast
be expected to be more extended than the main $qqq$
component. The implication is that the form factors may not
in the end be very sensitive to the range of the pionic
wave function component.

In the present phenomenological study, the probability of the
$qqqq\bar q$ component was determined by a fit to the empirical
electric form factor of the neutron. As this form factor obtains
no contribution from the $qqq$ component (unless the very small
contribution from the Wigner rotations are taken into account
\cite{Bruno}), it is exceptionally sensitive to minor wave function
components. There are several ways of providing a phenomenological
description of this form factor. To these belong inclusion of
a minor mixed symmetry $S-$state component and - as considered
here - explicit sea-quark components. The reality is most likely
to be a combination of several such mechanisms. The present
finding that the electric form factor of the neutron
can be described with a $qqqq\bar q$ component with $\sim$ 3\%
probability is in any case physically intuitive.

In the present study the wave function parameters and the
constituent mass were determined by a fit to $G_E^n$ and
the proton magnetic moment. With these parameters the
qualitative description of the momentum dependence of the
charge form factors of both the nucleons was achieved, as
well as of the magnetic form factors. While the 
calculated electric radii of the nucleon and the proton magnetic 
moment are close to the corresponding empirical values,
the calculated magnetic moment of the neutron is too
large. A more realistic value would need a combination
of such $qqqq\bar q$ configurations, which lead to the
same magnetic moment ratio for the nucleons as the
conventional $qqq$ wave functions. 

The present results represent a non-exclusive alternative to
the inclusion of mixed symmetry $S-$ and $D-$state components in the
wave function, which have previously been suggested as
partial explanations for the electric form factor of the
neutron \cite{Bruno}. As the presence of such wave function
components are a natural consequence of spin-dependent
hyperfine interactions between the constituent quarks,
the best description may most likely be obtained by
a combination of such effects with
sea-quark components of the form considered here.

%
%

\end{document}